# Cluster expansion approach to the effective potential in $\Phi^4_{2+1}$-theory*†

A. Peter, J.M. Häuser, M.H. Thoma, and W. Cassing

Institut für Theoretische Physik, Universität Giessen
35392 Giessen, Germany

February 16, 1995

**Abstract**

We apply a truncated set of dynamical equations of motion for connected equal-time Green functions up to the 4-point level to the investigation of spontaneous ground state symmetry breaking in $\Phi^4_{2+1}$ quantum field theory. Within our momentum space discretization we obtain a second order phase transition as soon as the connected 3-point function is included. However, an additional inclusion of the connected 4-point function still shows a significant influence on the shape of the effective potential and the critical coupling.

## 1 Introduction

In the last years the scalar quantum field theory with $\lambda\Phi^4$ self-interaction has become an important theoretical laboratory for probing the power of non-perturbative methods before applying them to more involved theories as e.g. QCD. This is because $\Phi^4$-theory allows to study different steps of the renormalization procedure by going from (1+1) [1]-[8], over (2+1) [6]-[10] to (3+1) space-time dimensions [6, 7, 10]-[13].

In a previous paper [14] we investigated spontaneous ground state symmetry breaking in the $\Phi^4$-theory in (1+1) dimensions by means of a closed set of equations of motion for connected equal-time Green functions which was derived from the coupled equations of motion for the full Green functions. The necessary truncation of the originally infinite set of equations was achieved by using the cluster expansions for n-point Green functions, i.e. their decomposition into sums of products of connected Green functions (correlation functions). Within this approach we found that an inclusion of the 3-point function is crucial for obtaining a second order phase transition from the symmetric ($\langle\Phi\rangle = 0$) to the symmetry broken ($\langle\Phi\rangle \neq 0$) phase. This has to be the case according to the Simon-Griffith theorem [15] and is in line with earlier investigations going beyond the Gaussian effective potential (GEP) [2]-[5]. Though the method was quite successful in its application to the (1+1)-dimensional theory it is not obvious what happens in (2+1) dimensions – which is the subject of the present paper – where we have to deal with a different topological structure[1] leading to an additional mass counterterm in the renormalization procedure.

*supported by DFG, BMFT, KFA Jülich and GSI Darmstadt
†part of the dissertation of A. Peter

The paper is organized as follows: In section 2 we introduce the connected Green function approach to the $\Phi^4$-theory in (2+1) dimensions. Section 3 contains the evaluation of the effective potential at zero temperature within various limits while section 4 is devoted to a summary. The complete set of dynamical equations in configuration space is presented in the appendix.

# 2 Correlation dynamics for $\Phi^4_{2+1}$-theory

The Lagrangian density and the corresponding Hamiltonian are given by

$$\mathcal{L} = \frac{1}{2}\partial_\mu\phi\partial^\mu\phi - \frac{1}{2}m_0^2\phi^2 - \frac{1}{4}\lambda\phi^4, \tag{1}$$

$$H = \frac{1}{2}\int d^2x \left[\pi^2 + (\vec{\nabla}\phi)^2 + m_0^2\phi^2 + \frac{1}{2}\lambda\phi^4\right], \tag{2}$$

respectively, with $\pi = \partial_t\phi$ and $m_0^2 = m^2 + \delta m^2$, where $m_0$ is the bare mass, $m$ the renormalized mass, and $\delta m^2$ is the mass counterterm. As in the (1+1) dimensional case the $\Phi^4_{2+1}$-theory is superrenormalizable and no coupling or wave function renormalization occurs. However, in addition to the divergent tadpole graph (fig.1a) in (1+1) dimensions, in (2+1) dimensions we have to subtract the divergent contribution due to the setting sun diagram (fig.1b). The corresponding divergent self-energies are given by

$$\Sigma^{tadpole}(m^2) = 3\lambda \int \frac{d^3p}{(2\pi)^3}\left[\frac{i}{p^2 - m^2 + i\epsilon}\right], \tag{3}$$

$$\Sigma^{sunset}(k^2, m^2) = -6i\lambda^2 \int \frac{d^3p}{(2\pi)^3}\int \frac{d^3q}{(2\pi)^3}\left[\frac{i}{p^2 - m^2 + i\epsilon}\right]\left[\frac{i}{q^2 - m^2 + i\epsilon}\right] \times \left[\frac{i}{(k-p-q)^2 - m^2 + i\epsilon}\right]. \tag{4}$$

[1] In the symmetry broken phase the two minima of the effective potential are unitary equivalent in $\Phi^4_{1+1}$-theory (which can be seen by the non-vanishing vacuum tunneling amplitude), while this is not the case in (2+1) dimensions (see e.g. [6, 16]).

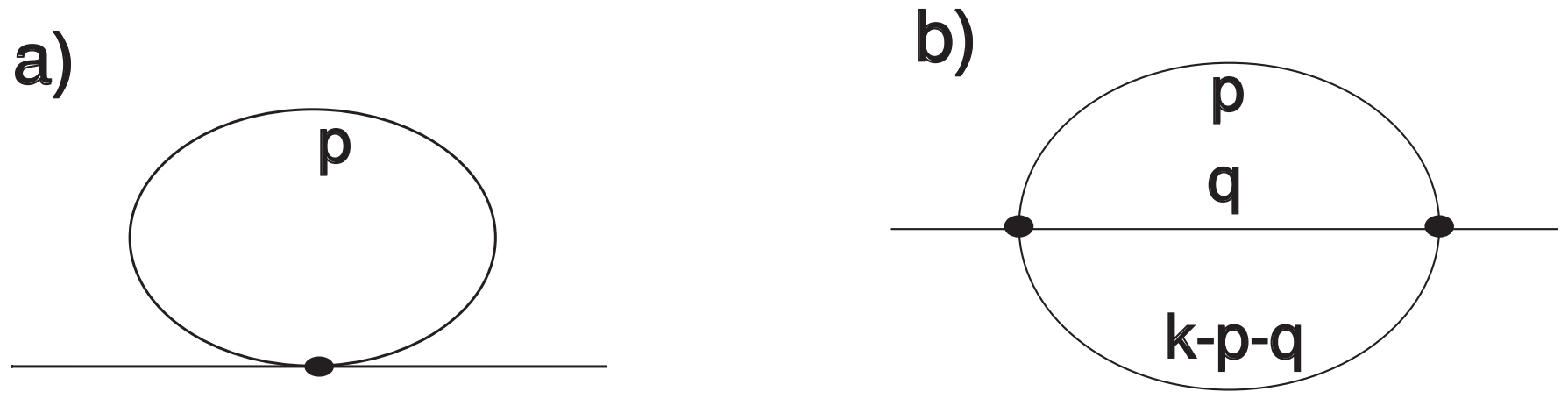

Figure 1: The primitively divergent self-energy graphs in the $\Phi^4_{2+1}$ field theory; tadpole (a), setting sun (b).

$\Sigma^{tadpole}(m^2)$ is linearly divergent and has the peculiarity of being independent of the external momentum, while the logarithmically divergent part of $\Sigma^{sunset}(k^2,m^2)$ is given by the first term of the corresponding Taylor expansion in the external momentum $k^2$ around $k^2=0$:

$$\begin{aligned}\Sigma^{sunset}(k^2,m^2) &= \sum_{n=0}^{\infty}\left(\frac{\partial^n}{\partial^n k^2}\Sigma^{sunset}(k^2,m^2)\mid_{k^2=0}\right)\frac{(k^2)^n}{n!}\\ &= \Sigma^{sunset}(0,m^2)+\tilde{\Sigma}^{sunset}(k^2,m^2).\end{aligned} \tag{5}$$

Since $\tilde{\Sigma}^{sunset}(k^2,m^2)$ is finite with $\tilde{\Sigma}^{sunset}(0,m^2)=0$ the mass counterterm in the zero-momentum-subtraction-scheme is given by

$$\delta m^2 = \delta m^2_{tadpole}+\delta m^2_{sunset} = \Sigma^{tadpole}(m^2)+\Sigma^{sunset}(0,m^2). \tag{6}$$

The subtraction of divergencies defined in this way is completely equivalent to the familiar BPH or BPHZ[2] renormalization scheme [17, 18].

Before subtracting the isolated divergencies within the method applied here, we have to perform the integrations over $dp_0$ in $\delta m^2_{tadpole}$ and over $dp_0, dq_0$ in $\delta m^2_{sunset}$. The results are:

$$\delta m^2_{tadpole} = -3\lambda\int\frac{d^2p}{(2\pi)^2}\frac{1}{2\omega_p} = -3\lambda\Delta(0), \tag{7}$$

$$\begin{aligned}\delta m^2_{sunset} = 6\lambda^2\int\frac{d^2p}{(2\pi)^2}\int\frac{d^2q}{(2\pi)^2}&\left\{\frac{1}{2\omega_q 2\omega_p([\omega_p+\omega_q]^2-\omega^2_{p+q})}\right.\\ &\left.+\frac{1}{2\omega_p 2\omega_{p+q}([\omega_{p+q}-\omega_p]^2-\omega_q^2)}+\frac{1}{2\omega_q 2\omega_{p+q}([\omega_{p+q}+\omega_q]^2-\omega_p^2)}\right\},\end{aligned} \tag{8}$$

with

$$\Delta(\vec{x}_1-\vec{x}_2) = \int\frac{d^2p}{(2\pi)^2}\frac{1}{2\omega_p}e^{i\vec{p}(\vec{x}_1-\vec{x}_2)}, \qquad \omega_p = \sqrt{\vec{p}^2+m^2}. \tag{9}$$

[2] Bogoliubov-Parasiuk-Hepp-Zimmermann

With the divergencies specified as above, we can now derive the closed set of equations of motion for equal-time Green functions up to 4-point functions.

We start by decomposing the field $\phi$ into a classical and a quantum part according to

$$\phi = \Phi_0 + \Phi, \qquad \langle\Phi\rangle = 0, \tag{10}$$

where $\Phi_0$ is a real constant. After this separation the Hamiltonian is given by

$$H = \frac{1}{2}\int d^2x \Big[\Pi^2 + (\vec{\nabla}\Phi)^2 + 2(\Phi_0 m^2 + \Phi_0\delta m^2 + \lambda\Phi_0^3)\Phi + (m^2 + \delta m^2 + 3\lambda\Phi_0^2)\Phi^2$$
$$+2\lambda\Phi_0\Phi^3 + \frac{1}{2}\lambda\Phi^4 + (m^2+\delta m^2)\Phi_0^2 + \frac{1}{2}\lambda\Phi_0^4\Big], \tag{11}$$

with $\pi = \partial_t\phi = \partial_t\Phi = \Pi$. Due to the inclusion of the mass counterterms in (11) the set of equations of motion generated by this Hamiltonian is already properly renormalized since the infinite constant zero-point contributions (still contained in the Hamiltonian) do not enter the Heisenberg equation.

By means of the equal-time commutation relations

$$[\Phi(\vec{x},t),\Phi(\vec{y},t)] = [\Pi(\vec{x},t),\Pi(\vec{y},t)] = 0, \quad [\Phi(\vec{x},t),\Pi(\vec{y},t)] = i\delta^2(\vec{x}-\vec{y}), \tag{12}$$

and the Heisenberg equation of motion we obtain

$$\partial_t\Phi = \Pi,$$

$$\partial_t\Pi = -\Phi_0(m^2+\delta m^2) - \lambda\Phi_0^3 + (\vec{\nabla}^2 - m^2 - \delta m^2 - 3\lambda\Phi_0^2)\Phi - 3\lambda\Phi_0\Phi^2 - \lambda\Phi^3 \tag{13}$$

for the time evolution of $\Phi$ and $\Pi$. In the same way we are now able to derive equations of motion for equal-time operator products of $\Phi$ and $\Pi$ of the general form

$$\hat{G}_n(\vec{x}_1,...,\vec{x}_n;t) = \prod_n(\Phi,\Pi), \tag{14}$$

where n is the number of field operators and all time arguments are taken to be equal as indicated by the single time argument $t$ occurring in $\hat{G}_n$ (e.g. for $n=2$ we obtain equations of motion for the operator products $\Phi(\vec{x}_1,t)\Phi(\vec{x}_2,t)$, $\Pi(\vec{x}_1,t)\Phi(\vec{x}_2,t)$, and $\Pi(\vec{x}_1,t)\Pi(\vec{x}_2,t)$). After taking the expectation values of these equations we obtain an infinite hierarchy of equations of motion for equal-time Green functions

$$\partial_t G_n = \partial_t\langle\hat{G}_n\rangle = \frac{1}{i}\langle[\hat{G}_n, H]\rangle, \tag{15}$$

where the equations of motion for n-point functions couple to (n+1)- and (n+2)-point functions due to the $\Phi^3$ and $\Phi^4$ terms in the Hamiltonian (11). We note in passing that for the description of the time evolution of a system in a local field theory it is sufficient to consider the equal-time Green functions, as long as one considers all Green functions containing the fields as well as their conjugate momenta.

In order to allow for a practical application, as e.g. the evaluation of the effective potential, we have to truncate the infinite set of equations. This can be achieved using the cluster expansion of n-point Green functions, i.e. their decomposition into sums of products of connected Green functions. The explicit form of the cluster expansions can be derived from the representation of the theory in terms of generating functionals of full and connected Green functions (see e.g. [14, 18, 19]).

Due to the separation of the field into a classical and a quantum part according to (10) all 1-point functions in the cluster expansions can be omitted[3]. Since our aim is a closed set of equations on the 4-point level and the equations of motion for the 4-point functions couple to the 5- and 6-point functions, we have to specify the cluster expansions up to sixth order, i.e.

$$
\begin{aligned}
\langle 12\rangle &= \langle 12\rangle_c, \\
\langle 123\rangle &= \langle 123\rangle_c, \\
\langle 1234\rangle &= \langle 1234\rangle_c + (1+\sum_{n=3}^{4}\mathcal{P}_{2n})\langle 12\rangle_c\langle 34\rangle_c, \\
\langle 12345\rangle &= \langle 12345\rangle_c + (1+\sum_{n=3}^{5}[\mathcal{P}_{1n}+\mathcal{P}_{2n}] + \sum_{n=3}^{4}\sum_{j=n+1}^{5}\mathcal{P}_{1n}\mathcal{P}_{2j})\langle 12\rangle_c\langle 345\rangle_c, \\
\langle 123456\rangle &= \langle 123456\rangle_c + (1+\sum_{n=4}^{6}\sum_{j=1}^{3}\mathcal{P}_{jn})\langle 123\rangle_c\langle 456\rangle_c \\
&\quad + (1+\sum_{n=3}^{6}[\mathcal{P}_{1n}+\mathcal{P}_{2n}] + \sum_{n=3}^{5}\sum_{j=n+1}^{6}\mathcal{P}_{1n}\mathcal{P}_{2j})\langle 12\rangle_c\langle 3456\rangle_c \\
&\quad + (1+\sum_{n=1}^{4}\mathcal{P}_{5n})(1+\sum_{n=3}^{4}\mathcal{P}_{2n})\langle 12\rangle_c\langle 34\rangle_c\langle 56\rangle_c. \qquad (16)
\end{aligned}
$$

In the cluster expansions (16) the integer numbers between the brackets stand for field operators $\Phi(i)$ or conjugate momenta $\Pi(i)$ while $\mathcal{P}_{ij}$ is the two-body permutation operator. We note that any application of the $\mathcal{P}_{ij}$ operators has to ensure the original order of the field operators within the connected Green functions occurring on the r.h.s. of (16). For a representation of the cluster expansions including the 1-point functions the reader is referred to ref. [19].

[3]We get the same result by using the Hamiltonian (2) instead of (11) and the complete cluster expansions with $\langle\phi\rangle = \Phi_0$.

The cluster expansions are truncated by neglecting all connected n-point Green functions with $n > 4$, i.e. $\langle 12345\rangle_c$ and $\langle 123456\rangle_c$ in (16). Inserting the truncated cluster expansions into the hierarchy of equations of motion (15) up to the 4-point functions on the l.h.s and hence up to the 6-point functions on the r.h.s. of (15) leads to the desired closed set of non-linear coupled equations of motion for the connected Green functions, i.e. a system of correlation dynamical equations[4].

Due to the inclusion of the mass counterterms in (11) the resulting set of equations of motion up to the 4-point level is properly renormalized as mentioned before. However, normal ordering of all operator products occurring in the equations of motion with respect to the mode expansions containing the renormalized mass $m$ (perturbative vacuum)

$$\Phi(\vec{x}) = \int \frac{d^2p}{(2\pi)^2} \frac{1}{\sqrt{2\omega_p}} \left[a(\vec{p}) + a^\dagger(-\vec{p})\right] e^{i\vec{p}\vec{x}},$$

$$\Pi(\vec{x}) = -i \int \frac{d^2p}{(2\pi)^2} \sqrt{\frac{\omega_p}{2}} \left[a(\vec{p}) - a^\dagger(-\vec{p})\right] e^{i\vec{p}\vec{x}}, \qquad \omega_p = \sqrt{\vec{p}^2 + m^2}, \tag{17}$$

shows that additional terms only arise within the connected 2-point functions, i.e.

$$\langle \Phi(\vec{x}_1)\Phi(\vec{x}_2)\rangle_c = \langle :\Phi(\vec{x}_1)\Phi(\vec{x}_2):\rangle_c + \Delta(\vec{x}_1 - \vec{x}_2),$$

$$\langle \Pi(\vec{x}_1)\Pi(\vec{x}_2)\rangle_c = \langle :\Pi(\vec{x}_1)\Pi(\vec{x}_2):\rangle_c - (\vec{\nabla}^2_{\vec{x}_1} - m^2)\Delta(\vec{x}_1 - \vec{x}_2),$$

$$\langle \Pi(\vec{x}_1)\Phi(\vec{x}_2)\rangle_c = \langle :\Pi(\vec{x}_1)\Phi(\vec{x}_2):\rangle_c - \frac{i}{2}\delta(\vec{x}_1 - \vec{x}_2),$$

$$\langle \Phi(\vec{x}_1)\Pi(\vec{x}_2)\rangle_c = \langle :\Phi(\vec{x}_1)\Pi(\vec{x}_2):\rangle_c + \frac{i}{2}\delta(\vec{x}_1 - \vec{x}_2), \tag{18}$$

where $\Delta(\vec{x}_1 - \vec{x}_2)$ is given in (9). This can be seen by a detailed analysis taking into account the cluster expansions (16).

After inserting the normal ordered 2-point functions (18) into the equations of motion, all terms proportional to $\Delta(0)$ arising from the normal ordering of $\langle \Phi(\vec{x}_1)\Phi(\vec{x}_1)\rangle_c$ cancel out due to the counterterm $\delta m^2_{tadpole}$ (7). But contrary to the theory in (1+1) dimensions, where the tadpole is the only divergent diagram [1, 9, 14], we are left with the contributions from the counterterm $\delta m^2_{sunset}$ (8) which do not cancel in the normal ordered equations of motion[5]. The resulting equations of motion for the 2-point

[4]This lengthy but straightforward derivation can most easily be performed by means of computer algebra, as e.g. provided by systems as MATHEMATICA.

[5]Chang and Magruder [9] have shown that even in $\Phi^4_{2+1}$-theory one is able to construct a finite Hamiltonian by introducing a new normal ordering prescription which, however, has to depend on the interaction explicitly.

connected Green functions read:

$$\frac{d}{dt}\langle :\Phi(1)\Phi(2):\rangle_c = \langle :\Pi(1)\Phi(2):\rangle_c + \langle :\Phi(1)\Pi(2):\rangle_c, \tag{19}$$

$$\begin{aligned}\frac{d}{dt}\langle :\Pi(1)\Phi(2):\rangle_c &= \langle :\Pi(1)\Pi(2):\rangle_c + \tilde{t}(1)\langle :\Phi(1)\Phi(2):\rangle_c - (\delta m^2_{sunset} + 3\lambda\Phi_0^2)\Delta(12) \\ &-\lambda[3\langle :\Phi(1)\Phi(1):\rangle_c(\langle :\Phi(1)\Phi(2):\rangle_c + \Delta(12)) + \langle :\Phi^3(1)\Phi(2):\rangle_c] \\ &-3\lambda\Phi_0\langle :\Phi(1)\Phi(1)\Phi(2):\rangle_c,\end{aligned} \tag{20}$$

$$\begin{aligned}\frac{d}{dt}\langle :\Pi(1)\Pi(2):\rangle_c &= \tilde{t}(1)\langle :\Phi(1)\Pi(2):\rangle_c \\ &-\lambda[3\langle :\Phi(1)\Phi(1):\rangle_c\langle :\Phi(1)\Pi(2):\rangle_c + \langle :\Phi^3(1)\Pi(2):\rangle_c] \\ &-3\lambda\Phi_0\langle :\Phi(1)\Phi(1)\Pi(2):\rangle_c \\ &+\tilde{t}(2)\langle :\Pi(1)\Phi(2):\rangle_c \\ &-\lambda[3\langle :\Pi(1)\Phi(2):\rangle_c\langle :\Phi(2)\Phi(2):\rangle_c + \langle :\Pi(1)\Phi^3(2):\rangle_c] \\ &-3\lambda\Phi_0\langle :\Pi(1)\Phi(2)\Phi(2):\rangle_c,\end{aligned} \tag{21}$$

with the modified kinetic energy

$$\tilde{t}(i) = t(i) - \delta m^2_{sunset} = \vec{\nabla}_i^2 - m^2 - 3\lambda\Phi_0^2 - \delta m^2_{sunset}. \tag{22}$$

Although normal ordering only affects the connected 2-point functions, we have written out the normal ordering operation $:\cdot:$ in all connected Green functions in order to have a uniform notation. Due to their length the equations of motion for the connected 3- and 4-point functions are not explicitly stated here and presented in the appendix.

Finally we expand the renormalized set of correlation dynamical equations of motion in an arbitrary orthonormal single-particle basis $\{\psi_\alpha\}$, i.e.

$$\Phi(\vec{x}) = \sum_\alpha \Phi_\alpha\psi_\alpha(\vec{x}), \qquad \Pi(\vec{x}) = \sum_\alpha \Pi_\alpha\psi_\alpha(\vec{x}), \tag{23}$$

which simplifies their numerical integration and additionally allows to discretize $\delta m^2_{sunset}$ in a consistent way. In order to compactify the resulting equations we introduce the following abbreviations:

$$\langle\alpha|\lambda_1\lambda_2\rangle = \int d^2x\ \psi^*_\alpha(\vec{x})\psi_{\lambda_1}(\vec{x})\psi_{\lambda_2}(\vec{x})\ ,$$

$$\langle\alpha|\lambda_1\lambda_2\lambda_3\rangle = \int d^2x\ \psi^*_\alpha(\vec{x})\psi_{\lambda_1}(\vec{x})\psi_{\lambda_2}(\vec{x})\psi_{\lambda_3}(\vec{x})\ ,$$

$$\tilde{t}_{\alpha\beta} = \int d^2x\ \psi^*_\alpha(\vec{x})\left(\vec{\nabla}^2 - m^2 - 3\lambda\Phi_0^2 - \delta m^2_{sunset}\right)\psi_\beta(x)\ ,$$

$$U_{\alpha\beta} = -3\lambda \sum_{\lambda_1\lambda_2} \langle\alpha|\lambda_1\lambda_2\beta\rangle\langle : \Phi_{\lambda_1}\Phi_{\lambda_2} :\rangle_c \, . \tag{24}$$

The permutation operator interchanging the indices $\alpha$ and $\beta$ is denoted by $\mathcal{P}_{\alpha\beta}$. Again we only present the equations on the 2-point level while the 3-point and 4-point functions follow in a straight forward manner from the appendix.

$$\frac{d}{dt}\langle : \Phi_\alpha\Phi_\beta :\rangle_c = \langle : \Pi_\alpha\Phi_\beta :\rangle_c + \langle : \Phi_\alpha\Pi_\beta :\rangle_c \, , \tag{25}$$

$$\begin{aligned}\frac{d}{dt}\langle : \Pi_\alpha\Phi_\beta :\rangle_c &= \langle : \Pi_\alpha\Pi_\beta :\rangle_c \\ &+ \sum_\lambda(\tilde{t}_{\alpha\lambda} + U_{\alpha\lambda})\langle : \Phi_\lambda\Phi_\beta :\rangle_c + \sum_\lambda(U_{\alpha\lambda} - \delta m^2_{sunset}\delta_{\alpha\lambda} - 3\lambda\Phi_0^2\delta_{\alpha\lambda})\Delta_{\lambda\beta} \\ &- \lambda \sum_{\lambda_1\lambda_2\lambda_3} \langle\alpha|\lambda_1\lambda_2\lambda_3\rangle\langle : \Phi_{\lambda_1}\Phi_{\lambda_2}\Phi_{\lambda_3}\Phi_\beta :\rangle_c \\ &- 3\lambda\Phi_0 \sum_{\lambda_1\lambda_2} \langle\alpha|\lambda_1\lambda_2\rangle\langle : \Phi_{\lambda_1}\Phi_{\lambda_2}\Phi_\beta :\rangle_c \, ,\end{aligned} \tag{26}$$

$$\begin{aligned}\frac{d}{dt}\langle : \Pi_\alpha\Pi_\beta :\rangle_c &= (1 + \mathcal{P}_{\alpha\beta})\sum_\lambda(\tilde{t}_{\alpha\lambda} + U_{\alpha\lambda})\langle : \Phi_\lambda\Pi_\beta :\rangle_c \\ &- \lambda(1 + \mathcal{P}_{\alpha\beta}) \sum_{\lambda_1\lambda_2\lambda_3} \langle\alpha|\lambda_1\lambda_2\lambda_3\rangle\langle : \Phi_{\lambda_1}\Phi_{\lambda_2}\Phi_{\lambda_3}\Pi_\beta :\rangle_c \\ &- 3\lambda\Phi_0(1 + \mathcal{P}_{\alpha\beta}) \sum_{\lambda_1\lambda_2} \langle\alpha|\lambda_1\lambda_2\rangle\langle : \Phi_{\lambda_1}\Phi_{\lambda_2}\Pi_\beta :\rangle_c.\end{aligned} \tag{27}$$

The resulting set of equations of motion for the $\Phi^4_{2+1}$-theory will be denoted as $\Phi^4_{2+1}CD$ ($\Phi^4_{2+1}$ *Correlation Dynamics*) furtheron. It conserves the total energy and momentum as can be shown within the general way outlined in [19].

# 3 Numerical solution of the $\Phi^4_{2+1}CD$ equations

In this section we present the application of the $\Phi^4_{2+1}CD$ equations to the evaluation of the effective potential of the $\Phi^4_{2+1}$-theory at zero temperature, which corresponds to the determination of the ground state energy for a given magnetization $\Phi_0$. Such a task automatically includes an investigation of the spontaneous symmetry breaking from the symmetric ($\langle\phi\rangle = \Phi_0 = 0$) to the symmetry broken ($\langle\phi\rangle = \Phi_0 \neq 0$) phase, i.e. a spontaneous breakdown of the symmetry under the transformation $\phi \to -\phi$ as soon as the coupling exceeds a critical value.

Since we are interested not only in the effective potential generated by the full set of equations but additionally in the influence of the 2-, 3- and 4-point functions we introduce 4 limiting cases of the $\Phi^4_{2+1}CD$ equations, denoted by $\Phi^4_{2+1}CD(2)$, $\Phi^4_{2+1}CD(2,3)$, $\Phi^4_{2+1}CD(2,4)$ and $\Phi^4_{2+1}CD(2,3,4)$. The numbers in parentheses indicate the various n-point functions taken into account, e.g. in $\Phi^4_{2+1}CD(2,3)$ the 4-point functions are set equal to zero while $\Phi^4_{2+1}CD(2,3,4)$ corresponds to the full set of equations.

At this point we again have to come back to the renormalization, since we only obtain meaningful results if we subtract the divergencies inherent in the introduced limiting cases. The divergent setting sun graph can only be generated when considering the 4-point function while the divergent contribution due to the tadpole is included in all 4 limiting cases of the $\Phi^4_{2+1}CD$ equations. Thus the mass counterterm is simply given by $\delta m^2 = \delta m^2_{tadpole}$ (as in the case of the (1+1) dimensional theory) as long as we restrict our calculations to the limits $\Phi^4_{2+1}CD(2)$ and $\Phi^4_{2+1}CD(2,3)$, hence we have to neglect all terms containing $\delta m^2_{sunset}$ in the equations of motion derived in section 2. However, in the $\Phi^4_{2+1}CD(2,4)$ and $\Phi^4_{2+1}CD(2,3,4)$ approximations our mass counterterm consists of the divergent contributions of both, the tadpole and the setting sun graph.

## 3.1 Application to the effective potential

In order to solve the set of equations (25) - (27) and those for the 3- and 4-point functions numerically, we discretize the momentum space by choosing plane waves in a two-dimensional box with periodic boundary conditions as a single-particle basis, i.e. we consider the theory on a torus. In order to select an appropriate box size for a given renormalized mass we have to make sure that our results converge as a function of the number of single-particle states included. Due to restrictions concerning computer time and memory space we have to compromise between a good momentum space resolution and the number of single-particle states involved. For practical purposes we therefore use a boxlength of 20 fm and include the 29 lowest lying single-particle states for a renormalized mass of 10 MeV. The convergence of our numerical studies with respect to the number of single-particle states will be demonstrated below (section 3.2).

As discussed in [14] there is no easy access to the stationary solutions of the correlation dynamical equations. Hence we are only able to describe the propagation of the system in time for a given initial configuration. In order to evaluate the ground state equal-time Green functions within the $\Phi^4_{2+1}CD$ approximation we thus exploit the Gell-Mann and Low theorem [20] by using the method of adiabatically switching on the coupling as was already done in [14]. Starting with the trivial exact ground state configuration for $\lambda = 0$ and a given fixed value of the magnetization $\Phi_0$ as an initial condition, we are then able to obtain the ground state equal-time Green functions for finite $\lambda$ as long as we switch on the coupling slow enough to ensure that the system adiabatically follows the trajectory of the ground state to a good approximation. The time-dependence of the coupling is chosen to be linear with $\lambda/(4m) = \beta t$ and $\beta = const$, but actually the functional form of the time-dependence is not essential as long as $\beta$ is small enough to guarantee adiabaticity. The convergence of the method, depending on $\beta$, was discussed in detail in [14]; it works equally well in (2+1) dimensions so that we do not present the convergence of the method again. For all results shown in the following we have chosen $\beta = 0.05$ c/fm, which yields a quite good compromise between the convergence of the method and the necessary computation time. Furthermore, in the subsequent discussions, we have replaced the $\Phi^4_{2+1}CD(2)$ limit of the correlation dynamical equations by the GEP approximation, since this represents the corresponding stationary limit. In order to present our results we use dimensionless quantities, e.g. $\langle H/m^3\rangle/L^2$ for the energy density, $\lambda/(4m)$ for the coupling, and $\Phi_0\sqrt{m}$ for the magnetization.

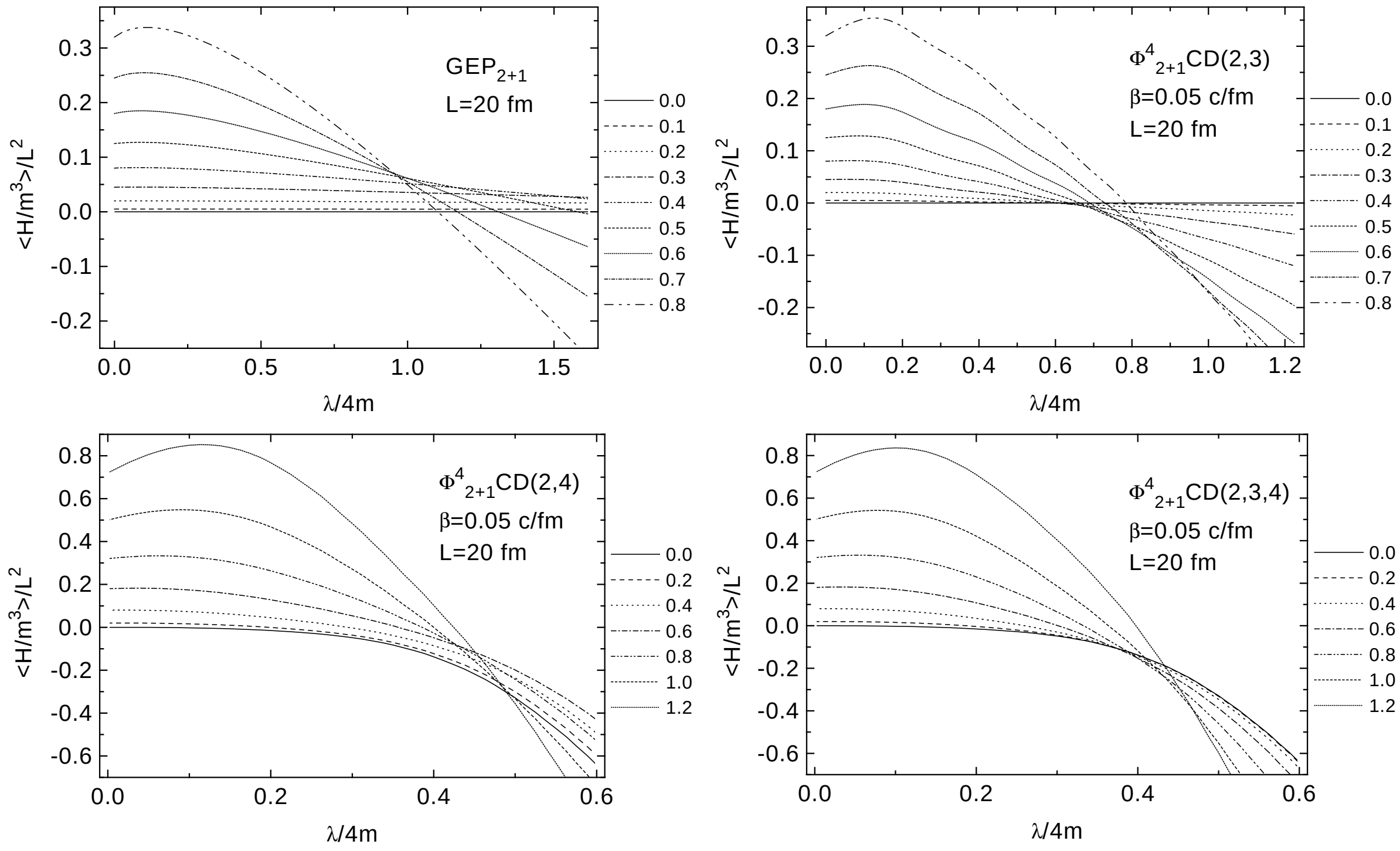

Figure 2: Ground state energy density as a function of the coupling for different values of the dimensionless vacuum magnetization; all curves (except for the GEP approximation) have been evaluated time-dependently with $\beta = 0.05$ c/fm.

Fig.2 shows the ground state energy density obtained as described above as a function of the coupling for various values of the magnetization. We observe that in all approximations there exists an intersection between the ground state energy density curve for $\Phi_0 = 0$ and a curve generated with a finite value of the magnetization, i.e. the energy minima are obtained for non-vanishing values of the magnetization when the coupling exceeds a critical value. Hence all limiting cases of the $\Phi^4_{2+1}CD$ equations predict a phase transition with increasing coupling. This result is in line with Magruder [9], who proves the existence of a phase transition in a properly renormal-ized (2+1) dimensional $\Phi^4$-theory by extending earlier arguments of Chang [1] for the (1+1) dimensional case. The proof is based on a comparison of two Lagrangians, one with positive and one with negative value of $m^2$, so that the corresponding classical potentials obtain their minimum for $\Phi_0 = 0$ and $\Phi_0 \neq 0$, respectively. One observes that a strong coupling theory defined by the Lagrangian with negative $m^2$ is equivalent to a weak coupling limit of the Lagrangian with positive $m^2$ and vice versa. Therefore the theory described by either of the two Lagrangians has to undergo a phase transition when the coupling is increased far enough. This concept of equivalent or dual Lagrangians was also exploited in [2] in case of the (1+1) dimensional $\Phi^4$-theory. According to this concept, which does not rely on any approximative method, the phase transition is governed by the renormalization, i.e. by the mass counterterms introduced above. We will come back to this point later and proceed by recognizing that we obtain a first order phase transition in case of the GEP and $\Phi^4_{2+1}CD(2,4)$ limits, while the $\Phi^4_{2+1}CD(2,3)$ and $\Phi^4_{2+1}CD(2,3,4)$ approximations lead to a phase transition of second order. This can seen by the fact that in case of the GEP and $\Phi^4_{2+1}CD(2,4)$ approach the curve with vanishing magnetization is first intersected by curves with considerably large values of the magnetization which implies that there is a discontinuity if we follow the trajectory of the lowest possible energy density, i.e. the ground state magnetization as a function of $\lambda$ has to jump from 0 to a finite value at the critical coupling. This is not the case for the $\Phi^4_{2+1}CD(2,3)$ and $\Phi^4_{2+1}CD(2,3,4)$ approximations, where all displayed curves with increasing magnetization intersect the curve with $\Phi_0 = 0$ in the correct order to yield a continuous increase of the ground state magnetization as a function of $\lambda$. But in contrast to the (1+1) dimensional theory we are not able to determine the proper order of the phase transition by simple considerations [1], since the Simon-Griffith theorem [15] is not applicable here.

In fig.3 the different effective potentials are plotted as a function of the magnetization $\Phi_0$ for some selected values of the coupling. The data in this figure are the same already used in the fig.2. In case of the GEP and $\Phi^4_{2+1}CD(2,4)$ approximations for couplings around the critical value there is an intermediate maximum between two minima located at $\Phi_0 = 0$ and at a finite value of the magnetization. The intermediate maximum leads to a sudden jump of the ground state magnetization from 0 to a finite value as soon as the critical value of the coupling is reached, yielding a phase transition of first order. In contrast to this the $\Phi^4_{2+1}CD(2,3)$ and $\Phi^4_{2+1}CD(2,3,4)$ approximations exhibit no intermediate maximum allowing for a continuous increase of the magnetization.

In addition to the order of the phase transition the different shapes of the effective potentials in fig.3 help to specify the regions where the 2-, 3- and 4-point functions reveal their main importance. Obviously, the 2-point function cannot lower the energy density around $\Phi_0 = 0$, while it becomes quite important for large values of $\Phi_0$. Due to the symmetry under the transformation $\phi \rightarrow -\phi$ the 3-point function has to vanish at $\Phi_0 = 0$, thus revealing its main importance at some intermediate value of $\Phi_0$. Within our $\Phi^4_{2+1}CD$ approximation the 4-point function is the only contribution leading to a reduction of the energy density around $\Phi_0 = 0$. These qualitative considerations illustrate why we obtain a phase transition of second order as soon as we include the 3-point function in our calculation. The 3-point function brings down the intermediate maximum (in the energy density) in a region where neither the 2-point nor the 4-point function can reduce the energy density substantially, yielding an effective potential characterized by only one minimum. A similar result concerning the relative importance of the different correlation functions was obtained in [14]. Therefore the qualitative picture that the n-point functions act dominantly at lower values of the magnetization than the (n–1)-point functions is confirmed again.

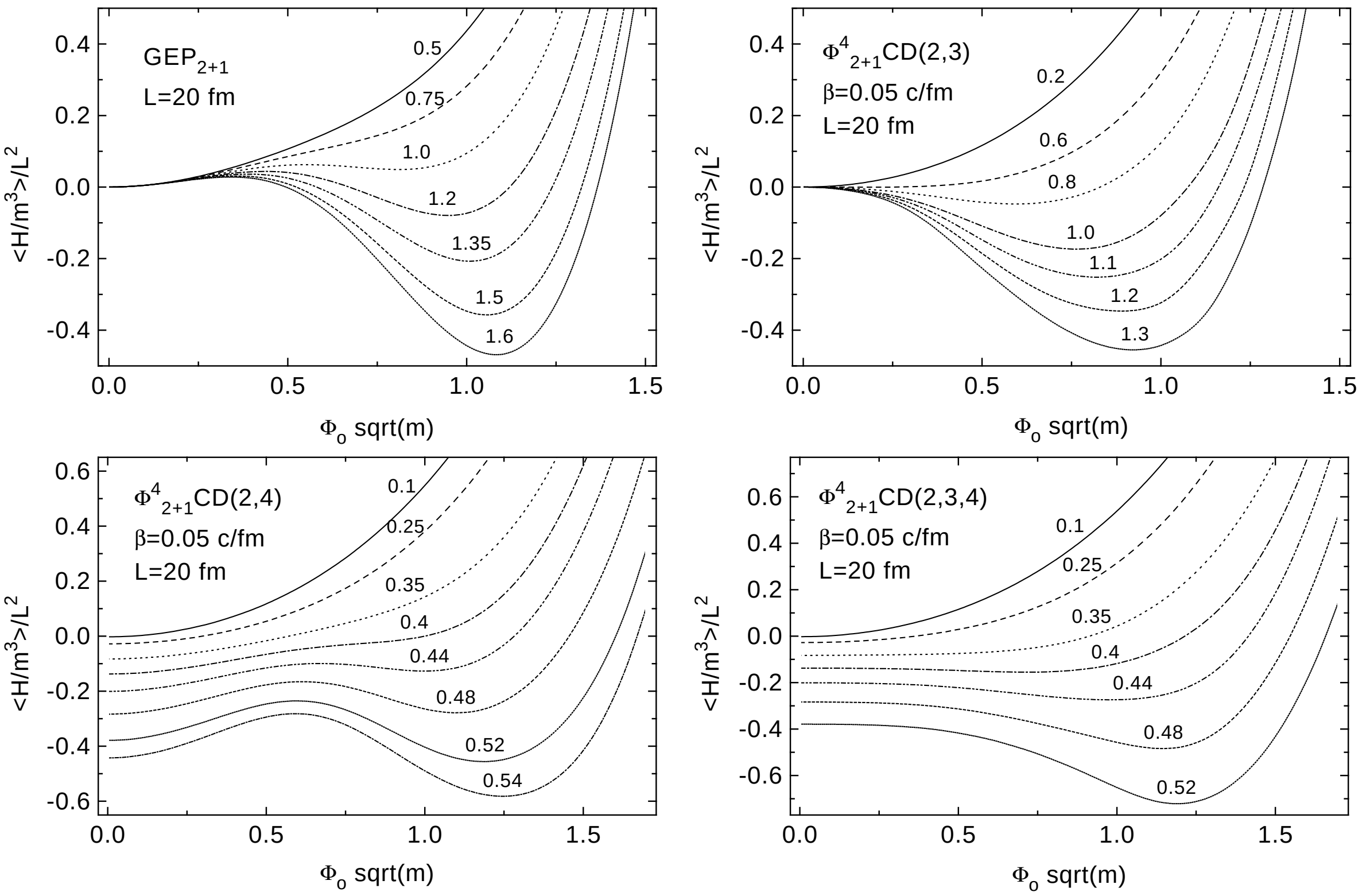

Figure 3: Effective potential as a function of the vacuum magnetization for different values of the dimensionless coupling $\lambda/(4m)$; all curves (except for the GEP approximation) have been evaluated time-dependently with $\beta = 0.05$ c/fm.

In order to present the phase transitions in a more common way the ground state magnetization (the order parameter of the phase transition) is plotted as a function of the coupling in fig.4. For this representation we again use the numerical data of figs. 2 and 3. In case of the GEP and $\Phi^4_{2+1}CD(2,4)$ approximations there is a discontinuity at the critical coupling yielding a sudden jump of the ground state magnetization from 0 to 0.89 and 1.1, respectively. Including the 3-point function in the $\Phi^4_{2+1}CD(2,3)$ and $\Phi^4_{2+1}CD(2,3,4)$ approximations we obtain a smooth increase of the ground state magnetization indicating a phase transition of second order.

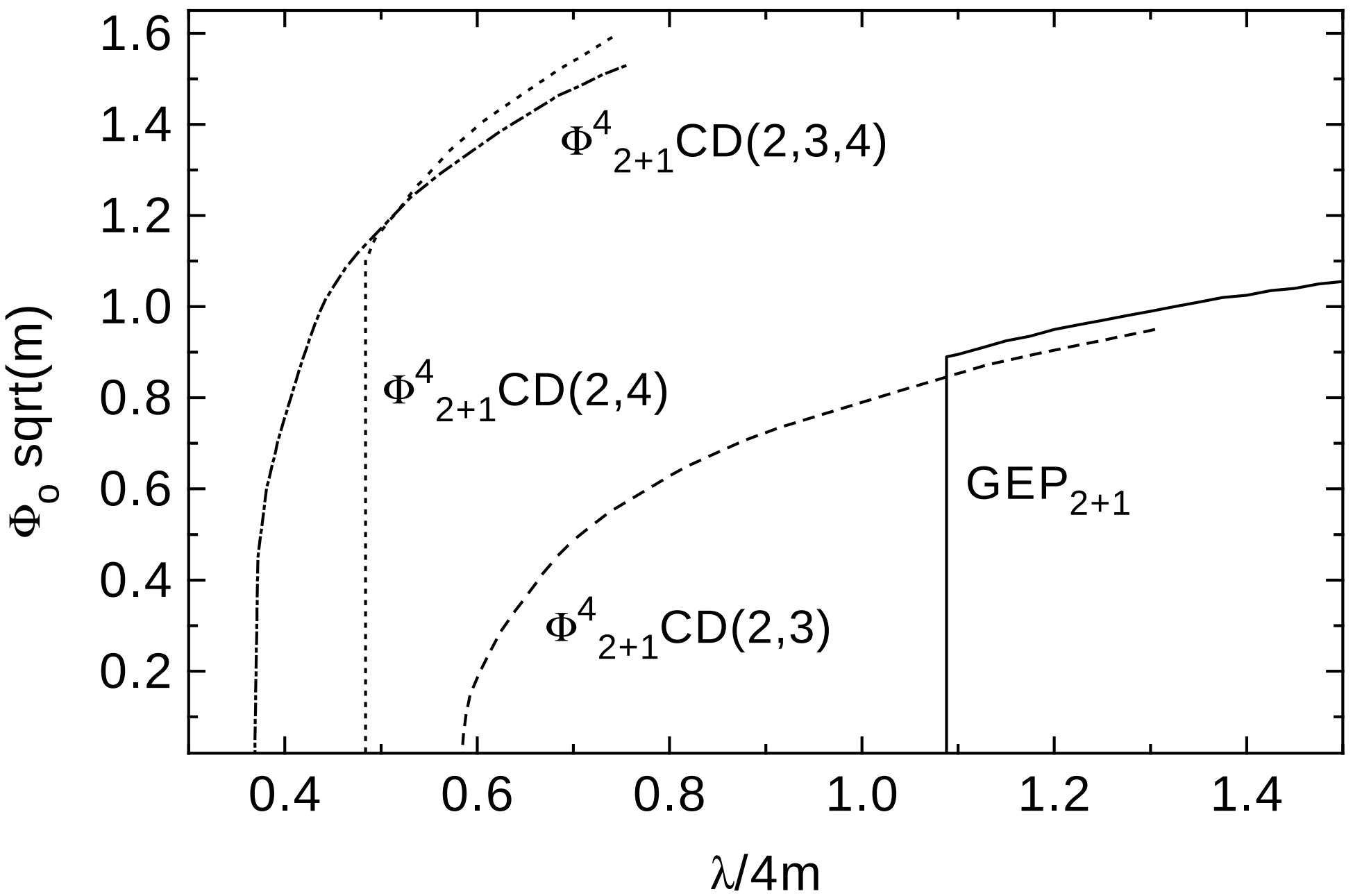

Figure 4: Ground state magnetization as a function of the dimensionless coupling for the different limiting cases of the $\Phi^4_{2+1}CD$ equations.

Furthermore, fig.4 shows that in the $\Phi^4_{2+1}CD(2,3,4)$ limit the phase transition is of 'weak' second order, meaning that the slope of the order parameter just above the critical coupling is too high in order to allow for a clear distinction from a first order transition. This observation can again easily be understood within our simple considerations concerning the shape of the effective potential. Since the 4-point function lowers the energy density around $\Phi_0 = 0$, the intermediate maximum, responsible for the first order phase transition in the GEP and $\Phi^4_{2+1}CD(2,4)$ limit, becomes more pronounced when the 4-point function is included.

Furthermore, fig.4 shows how the inclusion of the different n-point functions changes the critical coupling. The extracted numerical values are given in tab.1 in comparison to values obtained in [14] for the (1+1) dimensional $\Phi^4$-theory. Since the present investigation does not correspond to the continuum limit (see below) the actual quantitative values are not of primary interest, but a comparison of the relative values of the different approximations for the different dimensionalities allows to extract the influence of the divergent setting sun graph, which is only included in the $\Phi^4_{2+1}CD(2,4)$ and $\Phi^4_{2+1}CD(2,3,4)$ approximations. Obviously the inclusion of this diagram leads to a reduction of the critical coupling as can be seen by the fact that the $\Phi^4_{2+1}CD(2,4)$ and $\Phi^4_{2+1}CD(2,3,4)$ approximations have lower values of the critical coupling than the $\Phi^4_{2+1}CD(2,3)$ limit in contrast to the results obtained in the (1+1) dimensional theory.

| (1+1) dimensions | | (2+1) dimensions | |
|---|---|---|---|
| approximation | $(\lambda/4m^2)_{crit}$ | approximation | $(\lambda/4m)_{crit}$ |
| $GEP_{1+1}$ | 2.568 | $GEP_{2+1}$ | 1.088 |
| $\Phi^4_{1+1}CD(2,3)$ | 1.629 | $\Phi^4_{2+1}CD(2,3)$ | 0.585 |
| $\Phi^4_{1+1}CD(2,4)$ | 3.81 | $\Phi^4_{2+1}CD(2,4)$ | 0.484 |
| $\Phi^4_{1+1}CD(2,3,4)$ | 2.446 | $\Phi^4_{2+1}CD(2,3,4)$ | 0.369 |

Table 1: Values of the critical coupling for different approximations and space-time dimensions; the values in (1+1) dimensions, which are taken from [14], have been evaluated using $L$ = 100 fm boxlength, $\beta$ = 0.05 c/fm, $m$ = 10 MeV, and 15 single-particle states.

## 3.2 Numerical convergence

In order to get some feeling concerning the convergence of our results we show in fig.5 the critical coupling as a function of the momentum cutoff for different values of the boxsize for the GEP and $\Phi^4_{2+1}CD(2,3)$.

One clearly sees that the momentum discretization used for the presented results (L=20 fm boxlength, $(k/\Delta k)_{max} = 3$ which is equivalent to 29 single-particle states) is quite away from the continuum limit, where $(\lambda/4m)_{crit} = 3.078$ is the analytically known limit for the GEP [6] (straight line), but surely convergent with respect to the ultra-violet momentum cutoff. We furthermore observe that reducing the size of the momentum cell $\Delta k$ by increasing the boxlength (keeping the number of single-particle states fixed) yields higher values for the critical coupling while a higher value of the momentum cutoff – when increasing the number of basis states and keeping the size of the momentum cell fixed – has the opposite effect. Thus the analysis in the present work does not correspond to the continuum limit of the theory, but clearly demonstrates the relative importance of 2-, 3- and 4-point functions for the ground state energy density in different regimes of the coupling as well as for the order of the phase transition.

## 3.3 Comparison to related approaches

Finally we would like to discuss the results of our present work in the context of the literature known to us which also deals with the application of methods to the $\Phi^4$-theory in (2+1) dimensions, which go beyond the GEP approximation. Stancu and Stevenson

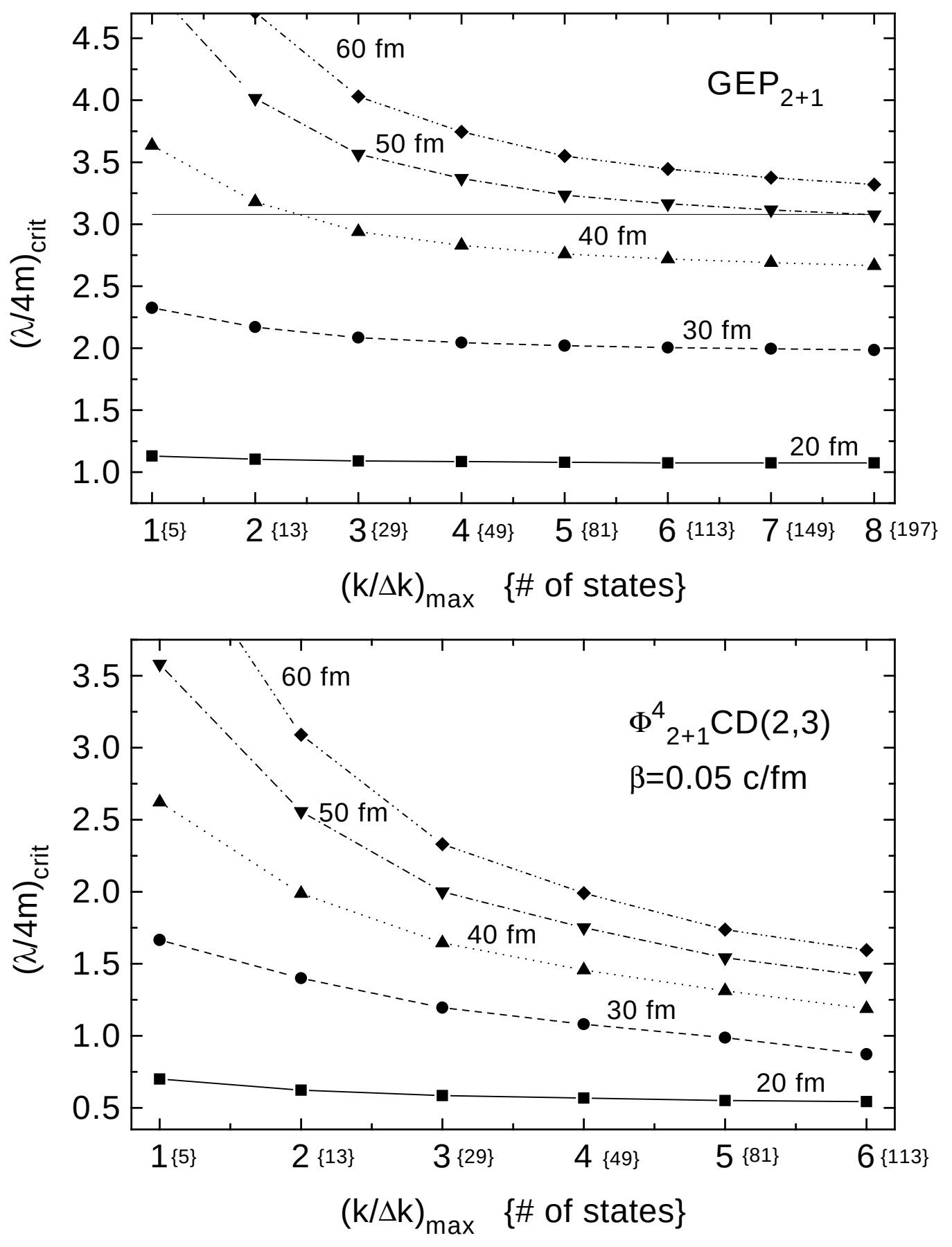

Figure 5: Critical coupling as a function of the momentum cutoff (i.e. the number of single-particle states) for different boxsizes; the values for the $\Phi^4_{2+1}CD(2,3)$ approximation have been evaluated time-dependently with $\beta = 0.05c/fm$; the horizontal line in the upper plot indicates the critical coupling for the GEP approximation in the continuum limit.

[7] have formulated a systematic non-perturbative expansion for the effective potential of the $\Phi^4$-theory based on the variational GEP method, thereby recovering, as a first order result, the GEP results. But in contrast to similar approaches as e.g. presented in [5, 8], the variational mass parameter is not fixed to the Gaussian value if one proceeds to higher orders of the expansion, which is, according to the authors, crucial for the expansion to yield convergent results. Going to the second order of this expansion (denoted by PGEP in [7]) yields a second order phase transition to the symmetry broken phase at a critical coupling of 2.775 or 2.237, for approaching the phase transition from below or above, respectively. In contrast to this result Cea and Tedesco [8] found that there is no spontaneous symmetry breaking in the (2+1) dimensional $\Phi^4$-theory by evaluating the two-loop corrections to the GEP approximation. This, however, is

in contradiction to our result and the proof given by Magruder [9]. The discrepancies arise due to the different use of the gap equation in [7] and [8], i.e. Stancu allows the variational mass parameter to change by going to the next order of the expansion while Cea keeps this value fixed. Such problems do not come up in the (1+1) dimensional theory where one only has the divergent tadpole graph which can consistently be subtracted in the GEP scheme yielding no ambiguities in the renormalization procedure when going beyond the GEP approach. In (2+1) dimensions difficulties occur when using the GEP as a starting point for a post Gaussian approach, since the GEP approximation takes into account only parts of the diagrams included in the second order of the loop expansion, e.g. the cactus graphs (see e.g. [1]), while the divergent setting sun diagram is not considered.

# 4 Summary

In this paper we have presented the application of a connected Green function approach (denoted by $\Phi^4 CD$) to the evaluation of the effective potential in (2+1) dimensional $\Phi^4$-theory and thereby to the investigation of the ground state symmetry breaking.

By comparing 4 limiting cases of the full $\Phi^4_{2+1}CD$ equations we were able to reconstruct where the 2-, 3- and 4-point functions dominantly influence the shape of the effective potential, which in turn allowed to give an intuitive explanation for the order of the phase transition obtained. The 2-point function mainly decreases the ground state energy for large values of the magnetization while the 3-point function exhibits its main influence at intermediate values. Both are not able to reduce the ground state energy density for vanishing magnetization ($\Phi_0 \to 0$) which, however, is the region where the 4-point function acts dominantly.

This picture provides a simple explanation of our results, showing a first order phase transition in the approximations which do not take into account the 3-point functions while the phase transition is of second order as soon as the 3-point function is included. The emerging picture is in line with our earlier results in [14] where the $\Phi^4 CD$ approach was applied to the (1+1) dimensional $\Phi^4$-theory.

In addition to the divergent tadpole graph of the $\Phi^4$-theory in (1+1) dimensions a proper renormalization in the (2+1) dimensional theory has to take into account the divergent setting sun graph, which is of second order in the coupling. Within our approximation schemes we had to care about this graph as soon as the 4-point function was included in our calculations, yielding a large reduction of the critical coupling. Therefore the $\Phi^4_{2+1}CD(2,4)$ and $\Phi^4_{2+1}CD(2,3,4)$ approximations lead to lower values of the critical coupling than the $\Phi^4_{2+1}CD(2,3)$ limit, in contrast to the results obtained in (1+1) dimensions. This clearly shows that the phase transition is governed by the counterterms taken into account. Thus the proper renormalization in each space-time dimensionality is essential for a correct description of the phase transition while the different approximations yield similar shapes for the effective potentials.

Since the description of the phase transition by the $\Phi^4CD$ equations leads to reasonable results in (1+1) and (2+1) dimensions and is in line with rigorous results known to us (e.g. the Simon-Griffith theorem [15]), we conclude, that correlation dynamics can be a powerful tool for the description of low-energy phenomena in local field theories. These quite encouraging results in the '$\Phi^4$-laboratory' thus seem to justify the application of the method to SU(N) gauge theories, aiming at a non-perturbative description of QCD phenomena. First steps in this direction have already been performed in [19].

# Appendix

## Equations of motion for the connected 3- and 4-point functions in configuration space

In this appendix we present the renormalized correlation dynamical equations for the normal ordered connected 3- and 4-point Green functions in configuration space in addition to (19)-(21) using the same notation:

$$\begin{aligned}\frac{d}{dt}\langle :\Phi(1)\Phi(2)\Phi(3):\rangle_c &= \langle :\Pi(1)\Phi(2)\Phi(3):\rangle_c \\ &+\langle :\Phi(1)\Pi(2)\Phi(3):\rangle_c + \langle :\Phi(1)\Phi(2)\Pi(3):\rangle_c,\end{aligned} \tag{28}$$

$$\begin{aligned}\frac{d}{dt}\langle :\Pi(1)\Phi(2)\Phi(3):\rangle_c &= \langle :\Pi(1)\Pi(2)\Phi(3):\rangle_c + \langle :\Pi(1)\Phi(2)\Pi(3):\rangle_c \\ &+\tilde{t}(1)\langle :\Phi(1)\Phi(2)\Phi(3):\rangle_c \\ &-3\lambda[\langle :\Phi(1)\Phi(1):\rangle_c\langle :\Phi(1)\Phi(2)\Phi(3):\rangle_c \\ &+\langle :\Phi(1)\Phi(1)\Phi(2):\rangle_c(\langle :\Phi(1)\Phi(3):\rangle_c + \Delta(13)) \\ &+(\langle :\Phi(1)\Phi(2):\rangle_c + \Delta(12))\langle :\Phi(1)\Phi(1)\Phi(3):\rangle_c] \\ &-3\lambda\Phi_0[2(\langle :\Phi(1)\Phi(2):\rangle_c + \Delta(12))(\langle :\Phi(1)\Phi(3):\rangle_c + \Delta(13)) \\ &+\langle :\Phi(1)\Phi(1)\Phi(2)\Phi(3):\rangle_c],\end{aligned} \tag{29}$$

$$\frac{d}{dt}\langle :\Pi(1)\Pi(2)\Phi(3):\rangle_c = \langle :\Pi(1)\Pi(2)\Pi(3):\rangle_c + (1+\mathcal{P}_{12})\tilde{t}(1)\langle :\Phi(1)\Pi(2)\Phi(3):\rangle_c$$
$$-3\lambda(1+\mathcal{P}_{12})[\langle :\Phi(1)\Phi(1):\rangle_c\langle :\Phi(1)\Pi(2)\Phi(3):\rangle_c$$
$$+\langle :\Phi(1)\Phi(1)\Pi(2):\rangle_c(\langle :\Phi(1)\Phi(3):\rangle_c + \Delta(13))$$
$$+\langle :\Phi(1)\Pi(2):\rangle_c\langle :\Phi(1)\Phi(1)\Phi(3):\rangle_c]$$
$$-3\lambda\Phi_0(1+\mathcal{P}_{12})[2\langle :\Phi(1)\Pi(2):\rangle_c(\langle :\Phi(1)\Phi(3):\rangle_c + \Delta(13))$$
$$+\langle :\Phi(1)\Phi(1)\Pi(2)\Phi(3):\rangle_c], \tag{30}$$

$$\frac{d}{dt}\langle :\Pi(1)\Pi(2)\Pi(3):\rangle_c = (1+\mathcal{P}_{12}+\mathcal{P}_{13})\tilde{t}(1)\langle :\Phi(1)\Pi(2)\Pi(3):\rangle_c$$
$$-3\lambda(1+\mathcal{P}_{12}+\mathcal{P}_{13})[\langle :\Phi(1)\Phi(1):\rangle_c\langle :\Phi(1)\Pi(2)\Pi(3):\rangle_c$$
$$+\langle :\Phi(1)\Phi(1)\Pi(2):\rangle_c\langle :\Phi(1)\Pi(3):\rangle_c$$
$$+\langle :\Phi(1)\Pi(2):\rangle_c\langle :\Phi(1)\Phi(1)\Pi(3):\rangle_c]$$
$$-3\lambda\Phi_0(1+\mathcal{P}_{12}+\mathcal{P}_{13})[2\langle :\Phi(1)\Pi(2):\rangle_c\langle :\Phi(1)\Pi(3):\rangle_c + \langle :\Phi(1)\Phi(1)\Pi(2)\Pi(3):\rangle_c]$$
$$+\frac{3}{2}\lambda\Phi_0[\delta(1-2)\delta(1-3) - \delta(1-2)\delta(2-3) + \delta(1-3)\delta(2-3)], \tag{31}$$

$$\frac{d}{dt}\langle :\Phi(1)\Phi(2)\Phi(3)\Phi(4):\rangle_c = \langle :\Pi(1)\Phi(2)\Phi(3)\Phi(4):\rangle_c$$
$$+\langle :\Phi(1)\Pi(2)\Phi(3)\Phi(4):\rangle_c + \langle :\Phi(1)\Phi(2)\Pi(3)\Phi(4):\rangle_c$$
$$+\langle :\Phi(1)\Phi(2)\Phi(3)\Pi(4):\rangle_c, \tag{32}$$

$$\frac{d}{dt}\langle :\Pi(1)\Phi(2)\Phi(3)\Phi(4):\rangle_c = \langle :\Pi(1)\Pi(2)\Phi(3)\Phi(4):\rangle_c$$
$$+\langle :\Pi(1)\Phi(2)\Pi(3)\Phi(4):\rangle_c + \langle :\Pi(1)\Phi(2)\Phi(3)\Pi(4):\rangle_c$$
$$+\tilde{t}(1)\langle :\Phi(1)\Phi(2)\Phi(3)\Phi(4):\rangle_c$$
$$-3\lambda[2(\langle :\Phi(1)\Phi(2):\rangle_c + \Delta(12))(\langle :\Phi(1)\Phi(3):\rangle_c + \Delta(13))(\langle :\Phi(1)\Phi(4):\rangle_c + \Delta(14))$$
$$+\langle :\Phi(1)\Phi(1)\Phi(2)\Phi(3):\rangle_c(\langle :\Phi(1)\Phi(4):\rangle_c + \Delta(14))$$
$$+\langle :\Phi(1)\Phi(1)\Phi(2)\Phi(4):\rangle_c(\langle :\Phi(1)\Phi(3):\rangle_c + \Delta(13))$$
$$+\langle :\Phi(1)\Phi(1)\Phi(2):\rangle_c\langle :\Phi(1)\Phi(3)\Phi(4):\rangle_c$$
$$+\langle :\Phi(1)\Phi(1)\Phi(3)\Phi(4):\rangle_c(\langle :\Phi(1)\Phi(2):\rangle_c + \Delta(12))$$
$$+\langle :\Phi(1)\Phi(1)\Phi(3):\rangle_c\langle :\Phi(1)\Phi(2)\Phi(4):\rangle_c + \langle :\Phi(1)\Phi(1)\Phi(4):\rangle_c\langle :\Phi(1)\Phi(2)\Phi(3):\rangle_c$$
$$+\langle :\Phi(1)\Phi(1):\rangle_c\langle :\Phi(1)\Phi(2)\Phi(3)\Phi(4):\rangle_c]$$
$$-6\lambda\Phi_0[(\langle :\Phi(1)\Phi(2):\rangle_c + \Delta(12))\langle :\Phi(1)\Phi(3)\Phi(4):\rangle_c$$
$$+(\langle :\Phi(1)\Phi(3):\rangle_c + \Delta(13))\langle :\Phi(1)\Phi(2)\Phi(4):\rangle_c$$
$$+(\langle :\Phi(1)\Phi(4):\rangle_c + \Delta(14))\langle :\Phi(1)\Phi(2)\Phi(3):\rangle_c], \tag{33}$$

$$
\begin{aligned}
\frac{d}{dt}\langle :\Pi(1)\Pi(2)\Phi(3)\Phi(4):\rangle_c &= \langle :\Pi(1)\Pi(2)\Pi(3)\Phi(4):\rangle_c + \langle :\Pi(1)\Pi(2)\Phi(3)\Pi(4):\rangle_c \\
&+(1+\mathcal{P}_{12})\tilde{t}(1)\langle :\Phi(1)\Pi(2)\Phi(3)\Phi(4):\rangle_c \\
&-3\lambda(1+\mathcal{P}_{12})[2\langle :\Phi(1)\Pi(2):\rangle_c(\langle :\Phi(1)\Phi(3):\rangle_c+\Delta(13))(\langle :\Phi(1)\Phi(4):\rangle_c+\Delta(14)) \\
&+\langle :\Phi(1)\Phi(1)\Pi(2)\Phi(3):\rangle_c(\langle :\Phi(1)\Phi(4):\rangle_c+\Delta(14)) \\
&+\langle :\Phi(1)\Phi(1)\Pi(2)\Phi(4):\rangle_c(\langle :\Phi(1)\Phi(3):\rangle_c+\Delta(13)) \\
&+\langle :\Phi(1)\Phi(1)\Pi(2):\rangle_c\langle :\Phi(1)\Phi(3)\Phi(4):\rangle_c + \langle :\Phi(1)\Phi(1)\Phi(3)\Phi(4):\rangle_c\langle :\Phi(1)\Pi(2):\rangle_c \\
&+\langle :\Phi(1)\Phi(1)\Phi(3):\rangle_c\langle :\Phi(1)\Pi(2)\Phi(4):\rangle_c + \langle :\Phi(1)\Phi(1)\Phi(4):\rangle_c\langle :\Phi(1)\Pi(2)\Phi(3):\rangle_c \\
&+\langle :\Phi(1)\Phi(1):\rangle_c\langle :\Phi(1)\Pi(2)\Phi(3)\Phi(4):\rangle_c] \\
&-6\lambda\Phi_0(1+\mathcal{P}_{12})[\langle :\Phi(1)\Pi(2):\rangle_c\langle :\Phi(1)\Phi(3)\Phi(4):\rangle_c \\
&+(\langle :\Phi(1)\Phi(3):\rangle_c+\Delta(13))\langle :\Phi(1)\Pi(2)\Phi(4):\rangle_c \\
&+(\langle :\Phi(1)\Phi(4):\rangle_c+\Delta(14))\langle :\Phi(1)\Pi(2)\Phi(3):\rangle_c], \qquad (34)
\end{aligned}
$$

$$
\begin{aligned}
\frac{d}{dt}\langle :\Pi(1)\Pi(2)\Pi(3)\Phi(4):\rangle_c &= \langle :\Pi(1)\Pi(2)\Pi(3)\Pi(4):\rangle_c \\
&+(1+\mathcal{P}_{12}+\mathcal{P}_{13})\tilde{t}(1)\langle :\Phi(1)\Pi(2)\Pi(3)\Phi(4):\rangle_c \\
&-3\lambda(1+\mathcal{P}_{12}+\mathcal{P}_{13})[2\langle :\Phi(1)\Pi(2):\rangle_c\langle :\Phi(1)\Pi(3):\rangle_c(\langle :\Phi(1)\Phi(4):\rangle_c+\Delta(14)) \\
&+\langle :\Phi(1)\Phi(1)\Pi(2)\Pi(3):\rangle_c(\langle :\Phi(1)\Phi(4):\rangle_c+\Delta(14)) \\
&+\langle :\Phi(1)\Phi(1)\Pi(2)\Phi(4):\rangle_c\langle :\Phi(1)\Pi(3):\rangle_c + \langle :\Phi(1)\Phi(1)\Pi(2):\rangle_c\langle :\Phi(1)\Pi(3)\Phi(4):\rangle_c \\
&+\langle :\Phi(1)\Phi(1)\Pi(3)\Phi(4):\rangle_c\langle :\Phi(1)\Pi(2):\rangle_c + \langle :\Phi(1)\Phi(1)\Pi(3):\rangle_c\langle :\Phi(1)\Pi(2)\Phi(4):\rangle_c \\
&+\langle :\Phi(1)\Phi(1)\Phi(4):\rangle_c\langle :\Phi(1)\Pi(2)\Pi(3):\rangle_c \\
&+\langle :\Phi(1)\Phi(1):\rangle_c\langle :\Phi(1)\Pi(2)\Pi(3)\Phi(4):\rangle_c] \\
&-6\lambda\Phi_0(1+\mathcal{P}_{12}+\mathcal{P}_{13})[\langle :\Phi(1)\Pi(2):\rangle_c\langle :\Phi(1)\Pi(3)\Phi(4):\rangle_c \\
&+\langle :\Phi(1)\Pi(3):\rangle_c\langle :\Phi(1)\Pi(2)\Phi(4):\rangle_c + (\langle :\Phi(1)\Phi(4):\rangle_c+\Delta(14))\langle :\Phi(1)\Pi(2)\Pi(3):\rangle_c] \\
&+\frac{3}{2}\lambda\delta(1-3)\delta(2-3)[\langle :\Phi(3)\Phi(4):\rangle_c+\Delta(34)], \qquad (35)
\end{aligned}
$$

$$
\begin{aligned}
\frac{d}{dt}\langle :\Pi(1)\Pi(2)\Pi(3)\Pi(4):\rangle_c &= (1+\mathcal{P}_{12}+\mathcal{P}_{13}+\mathcal{P}_{14})\tilde{t}(1)\langle :\Phi(1)\Pi(2)\Pi(3)\Pi(4):\rangle_c \\
&-3\lambda(1+\mathcal{P}_{12}+\mathcal{P}_{13}+\mathcal{P}_{14})[2\langle :\Phi(1)\Pi(2):\rangle_c\langle :\Phi(1)\Pi(3):\rangle_c\langle :\Phi(1)\Pi(4):\rangle_c \\
&+\langle :\Phi(1)\Phi(1)\Pi(2)\Pi(3):\rangle_c\langle :\Phi(1)\Pi(4):\rangle_c + \langle :\Phi(1)\Phi(1)\Pi(2)\Pi(4):\rangle_c\langle :\Phi(1)\Pi(3):\rangle_c \\
&+\langle :\Phi(1)\Phi(1)\Pi(2):\rangle_c\langle :\Phi(1)\Pi(3)\Pi(4):\rangle_c + \langle :\Phi(1)\Phi(1)\Pi(3)\Pi(4):\rangle_c\langle :\Phi(1)\Pi(2):\rangle_c \\
&+\langle :\Phi(1)\Phi(1)\Pi(3):\rangle_c\langle :\Phi(1)\Pi(2)\Pi(4):\rangle_c + \langle :\Phi(1)\Phi(1)\Pi(4):\rangle_c\langle :\Phi(1)\Pi(2)\Pi(3):\rangle_c \\
&+\langle :\Phi(1)\Phi(1):\rangle_c\langle :\Phi(1)\Pi(2)\Pi(3)\Pi(4):\rangle_c] \\
&-6\lambda\Phi_0(1+\mathcal{P}_{12}+\mathcal{P}_{13}+\mathcal{P}_{14})[\langle :\Phi(1)\Pi(2):\rangle_c\langle :\Phi(1)\Pi(3)\Pi(4):\rangle_c \\
&+\langle :\Phi(1)\Pi(3):\rangle_c\langle :\Phi(1)\Pi(2)\Pi(4):\rangle_c + \langle :\Phi(1)\Pi(4):\rangle_c\langle :\Phi(1)\Pi(2)\Pi(3):\rangle_c]
\end{aligned}
$$

$$+\frac{3}{2}\lambda[\delta(1-3)\delta(2-3)\langle:\Phi(3)\Pi(4):\rangle_c+\delta(1-4)\delta(3-4)\langle:\Pi(2)\Phi(4):\rangle_c$$
$$+\delta(1-4)\delta(2-4)\langle:\Pi(3)\Phi(4):\rangle_c+\delta(2-4)\delta(3-4)\langle:\Pi(1)\Phi(4):\rangle_c]. \qquad (36)$$